%Paper: hep-th/9212039
%From: Kimura Kazuhiro <kimura@kurims.kyoto-u.ac.jp>
%Date: Sat, 5 Dec 92 15:27:39 JST

%%%%%%%%%%%%%%%%%%%%%%%%%%%%%%%%%%%%%%%%%%%%%%%%%%%%%
\baselineskip=19pt
\magnification=\magstep1
\font\germ=eufm10

\font\fontB=cmr12

\def\sl{{\hbox{\germ sl}}\,}
\def\ge{{\hbox{\germ g}}\,}

\def\and{\hbox{ \  and \ }}

\def\medn{\medskip\medskip\noindent}
\def\med{\medskip\medskip}

\def\vbn{\vfill\break\noindent}

\def\qb{q^{{k \over 2}m}}
\def\qa{q^{-{k \over 2}m}}
\def\qkb{q^{(k+2)m}}
\def\qka{q^{-(k+2)m}}
\def\qq{q-q^{-1}}

%%%%%%%%%%%%%%%%%%%%%%%%%%%%%%%%%%%%%%%%%%%%%%%%%%%%%
\footline={\hfill}
\vskip 2pt
{\rm \hfill
\medskip
{\fontB
\centerline{On Free Boson Representations of the Quantum Affine Algebra
$U_q(\widehat{\sl}_2)$}}
\medskip\medskip
\vskip 45pt
\centerline{Kazuhiro KIMURA}\med
\centerline{\it Research Institute for Mathematical Sciences}\par
\centerline{\it Kyoto University, Kyoto 606-01, Japan}
\med\medn
\vskip 20pt
\centerline{\rm December 1992}
\vskip 50pt
\centerline{Dedicated to Professor N. Nakanishi
on the occation of his sixtieth birthday}
\vskip 70pt
\centerline{\fontB{ ABSTRACT}}
\vskip 7pt
\noindent
A boson representation of the quantum affine algebra $U_q(\widehat{\sl}_2)$
is realized based on the Wakimoto construction. We discuss
relations with the other boson representations.
\vbn
\pageno=2
\footline{\hfill-\ \folio \ -\hfill}
\beginsection{1. Introduction}\par
It is established that  conformal field theories play an important role
in studies on two dimensional models.
 In particular the
Wess-Zumino-Witten(WZW) model provides us with powerful tools
to investigate models in the language of affine Kac-Moody algebras.
It is also recognized that free field realization is indeed useful to
investigate representation of Virasoro and affine Kac-Moody algebras.
Wakimoto[1] first introduced free realization of the affine Kac-Moody algebra
$\widehat{\sl}_2$ and there exist many works on this realization.
\par
Recently Frenkel and Reshetikhin[2] have constructed certain $q$-deformed
chiral vertex operators(qVOs) of the WZW model
based on the representation theory of the quantum affine algebra
$U_q(\hat\ge)$. They showed that the correlation functions satisfy
a $q$-difference equation called the $q$-deformed Knizhnik-Zamolodchikov(qKZ)
equation,
and the connection matrix of the qKZ equation
indeed corresponds with the elliptic solution
of the Yang-Baxter equation. As an application XXZ models
are analyzed by the technique of $q$-vertex operators[3][4][5].
\par
In these situations it is desirable to construct a concrete realization
of quantum affine algebras and $q$-deformed chiral vertex operators.
Frenkel and Jing first found a $q$-deformation of the Frenkel-Kac
construction which corresponds to  boson representation of
$U_q(\widehat{\sl}_2)$ of level one[6].
They show the Drinfeld realization[7] can be treated in terms of currents
in which the technique of operator product expansions(OPEs) is powerful.
Following this work Jimbo et al. introduced explicit forms of
$q$-deformed chiral vertex operators and calculated the trace of
the product of the vertex operators.
 Recently some papers appear
to attempt extending to the case of arbitrary level. There are two
kinds of boson realization in these papers. One is based
on the Wakimoto construction with a bosonization of bosonic ghost
system[8] and the other on the realization in terms of bosonized
parafermions[9][10].
In this letter we construct another boson representation $\grave {\rm a}$
la Wakimoto following the line of the Matsuo's realization[9].
In his formulation currents of  $U_q(\widehat{\sl}_2)$ split
into two parts of creation and annihilation, which is useful to
investigate the structure of the algebra.
We also show explicit relations between his realization and ours.
\beginsection{2.
Wakimoto Construction of the algebra $\widehat{\sl}_2$ }\par
\noindent
We begin with a brief review of the Wakimoto description of
the algebra $\widehat{\sl}_2$ and its bosonized representation.
In this construction the $\widehat{\sl}_2$ WZW model is described
in terms of one free boson $\varphi$ and a bosonic $(\beta,\gamma)$
system of weight (1,0). We can express the algebraic relations
in the language of OPEs:
$$
\beta(z)\gamma(w)={1 \over z-w}+:\beta(z)\gamma(w):,
$$
$$
\varphi(z)\varphi(w)=\ln(z-w)+:\varphi(z)\varphi(w):,
\eqno(1)
$$
where :: denotes normal ordering.
With the mode expansion:
$$
\beta(z)=\sum_{n\in Z} \beta_nz^{-(n+1)},
\qquad \gamma(z)=\sum_{n\in Z} \gamma_nz^{-n},
$$
$$
\varphi(z)=\alpha-\alpha_0\ln z+\sum_{n\ne0}{\alpha_n \over n}z^{-n},
\eqno(2)
$$
we have
$$
[\beta_m,\gamma_n]=\delta_{m+n,0},
$$
$$
[\alpha_0,\alpha]=1,\qquad [\alpha_m,\alpha_n]=m\delta_{m+n,0}.
\eqno(3)
$$
The currents[1] are given by these fields:
$$
\eqalign{
&J^+(z)=\beta(z), \cr
&J^-(z)=-:\beta(z)\gamma^2(z):+\sqrt{2(k+2)}\partial\varphi(z)\gamma(z)
+k\partial\gamma(z),  \cr
&J^3(z)=-:\beta(z)\gamma(z):+\sqrt{{k+2 \over 2}}\partial\varphi(z), \cr
}
\eqno(4)
$$
\noindent
where k is a level of the Kac-Moody algebra. OPEs
of these currents are derived from ones of the fields $\varphi,
(\beta,\gamma)$:
$$
\eqalign{
&J^3(z)J^3(w) \sim {k \over (z-w)^2}, \cr
&J^3(z)J^{\pm}(w) \sim \pm{ 1 \over z-w}J^\pm(w),
\qquad\vert z\vert>\vert w\vert, \cr
&J^+(z)J_-(w)\sim{k \over (z-w)^2}+{1 \over z-w}J^3(w),  \cr
}
\eqno(5)
$$
\noindent
which is equivalent to the commutation relations:
$$
\eqalign{
&[J^3_m,J^3_n]=kn\delta_{m+n,0}, \cr
&[J^3_m,J^\pm_n]=J^\pm_{m+n},  \cr
&[J^+_m,J^-_n]=J^3_{m+n}+km\delta_{m+n,0}, \cr
}
\eqno(6)
$$
\noindent
where
$$
H(z)=\sum_{n\in Z}J^3_nz^{-(n+1)},
\qquad J^{\pm}(z)=\sum_{n\in Z} J^\pm_nz^{-(n+1)}.
$$
\noindent
The primary fields of the spin $j$ representation and
screening charge are written as
$$
\phi_{j,m}(z)=\{\gamma(z)\}^{j-m}:\exp\biggl\{\sqrt{{2 \over k+2}}j
\varphi(z)\biggr\}:,
\qquad 0\le j\le k,\quad -j\le m\le j,
$$
$$
J^3(z)\phi_{j,m}(w)\sim {m \over z-w}\phi_{j,m}(w).
$$
$$
J^S(z)=-\beta(z):\exp\biggl\{-\sqrt{{2 \over k+2}}\varphi(z)\biggr\}:.
\eqno(7)
$$
In order to construct $q$-deformation of the Wakimoto currents, it is
convenient to use the bosonized representation of the $(\beta,\gamma)$ system
[11]:
$$
\beta=-:\partial\chi(z)\exp\bigl\{-\chi(z)+i\sigma(z)\bigr\}:,\qquad \gamma(z)=
:\exp\bigl\{\chi(z)-i\sigma(z)\bigr\}:,
\eqno(8)
$$
\noindent
where $\chi(z)$ and $\sigma(z)$ are bosonic fields and OPEs of them
are same as those of
$\varphi$. With these fields the bosonized
Wakimoto currents[12] are rewritten as
$$
\eqalign{
&J^+(z)=-:\partial\chi(z)\exp\bigl\{-\chi(z)+i\sigma(z)\bigr\}:, \cr
&J^-(z)=:\bigl[(k+2)\partial\bigl\{\chi(z)-i\sigma(z)\bigr\}-\partial\chi(z)
+\sqrt{2(k+2)}\partial\varphi(z)\bigr]\exp\bigl\{\chi(z)-i\sigma(z)\bigr\}:,
\cr
&J^3(z)=-i\partial\sigma(z)+\sqrt{k+2 \over 2}\partial\varphi(z). \cr
}
\eqno(9)
$$
\noindent
The primary fields and screening charge now become
$$
\eqalign{
&\phi_{j,m}(z)=:\exp\biggl[(j-m)\bigl\{\chi(z)-i\sigma(z)\bigr\}\biggr]
\exp\biggl\{\sqrt{{2 \over k+2}}j\varphi(z)\biggr\}:, \cr
&J^S(z)=-:\partial\chi(z)\exp\bigl\{-\chi(z)+i\sigma(z)\bigr\}
\exp\biggl\{-{ \sqrt{2\over k+2}}\varphi(z)\biggr\}:. \cr
}
\eqno(10)
$$
\beginsection{3. Quantum Affine Algebra $U_q(\widehat{\sl}_2)$}\par
\noindent
It is known that the algebra $U_q(\sl_2)$  can be realized in terms of
a $q$-deformed harmanic oscillator. First we introduce $q$-deformed
oscillators of annihilation and creation, $\it a$ and $a^{\dagger}$,
which satisfy following relations:
$$
\eqalign{
&a a^{\dagger}-qa^{\dagger}a=q^{-N}, \cr
&a a^{\dagger}-q^{-1}a^{\dagger}a=q^{N}, \cr
&[N,a^{\dagger}]=a^{\dagger}, \cr
&[N,a]=-a, \cr
}
\eqno(11)
$$
\noindent
where $q$ is a deformation parameter.
Hamiltonian of a $q$-deformed harmonic oscillator is given by
${\cal H}=[N]+{1 \over 2}$, where we use the notation
$[N]={q^N-q^{-N} \over q-q^{-1}}$. One can obtain the $q$-deformed
$\sl(2)$algebra from these oscillators.
$$
\eqalign{
&J^+=a^{\dagger}, \cr
&J^-=a[\lambda+1-N],  \cr
&J^3=N-{\lambda \over 2}, \cr
}
\eqno(12)
$$
\noindent
where $\lambda$ is constant corresponding to a value of spin.
In the case of the irreducible representation $\lambda=j$($\it j$ integer),
weight vectors are given by
$$
\mid l>={(a^{\dagger})^l \over \sqrt{[l]!}}
\mid0>,\qquad
[N]\mid0>=0,\qquad l=0,1,2,\cdot\cdot\cdot,j.
\eqno(13)
$$
These are eigenvectors of the $q$-deformed Number operator[$\it N$]
belonging to eigenvalues $[l]$.
It is easy to check these operators satisfy the commutation relations of
the quantum $\sl(2)$ algebra:
$$
\eqalign{
&[J^3,J^\pm]=\pm J^\pm, \cr
&[J^+,J^-]=[2J^3]={q^{2J^3}-q^{-2J^3} \over q-q^{-1}}. \cr
}
\eqno(14)
$$
\par
Now we are in a position to construct the $q$-deformation of
the Wakimoto current in the same spirit as the  $\widehat{\sl}_2$
case. First we introduce $q$-deformation of the boson $\varphi(z)$:
$$
\varphi(z)=\alpha-\alpha_0\ln z+\sum_{n\ne0}{\alpha_n \over [n]}z^{-n},
$$
$$
[\alpha_0,\alpha]=2,
$$
$$
[\alpha_m,\alpha_n]=\delta_{m+n,0}{[2m][m] \over m}.
\eqno(15)
$$
In the case of a system with finite freedom, one has to use
$q$-deformed oscillators. However, in field theories
one can use ordinary oscillators
with adequate normalization. As the Wakimoto currents contain
three bosons $\{\sigma(z),\chi(z),\varphi(z)\}$, we prepare
three kinds of oscillators$\{a_n, \bar a_n, b_n;n\in Z;a,\bar a,b\}$
satisfying the following commutation relations:
$$
\eqalign{
&[a_m,a_n]=-\delta_{m+n,0}{[2m][2m]\over m},\qquad [a_0,a]=-4, \cr
&[\bar a_m,\bar a_n]=\delta_{m+n,0}{[2m][2m]\over m},
\qquad [\bar a_0,\bar a]=4, \cr
&[b_m,b_n]=\delta_{m+n,0}{[2m][(k+2)m]\over m},\qquad [b_0,b]
=2(k+2). \cr
}
\eqno(16)
$$
The other commutation relations of oscillators are equal to zero.
We define currents of the algebra $U_q(\widehat{\sl}_2)$ by using
the oscillators(16) as follows:
$$
\eqalign{
K_+(z)=&\exp\biggl\{(q-q^{-1})\sum_{m=1}^{\infty}z^{-m}(a_m+b_m)\biggr\}
q^{(a_0+b_0)},  \cr
K_-(z)=&\exp\biggl\{-(q-q^{-1})\sum_{m=1}^{\infty}z^{m}
(q^{-(k+2)m}a_{-m}+q^{-2m}b_{-m})\biggr\}
q^{-(a_0+b_0)},  \cr
X^+(z)=&{1 \over q-q^{-1}}:\biggl\{Y^+(z)Z_+(q^{-{k+2 \over 2}}z)
-Z_-(q^{{k+2 \over 2}}z)Y^+(z)\biggr\}:,    \cr
X^-(z)=&-{1 \over q-q^{-1}}:\biggl\{Y^-(z)Z_+(q^{{k+2 \over 2}}z)
U_+(q^{k \over 2}z)W_+(q^{k \over 2}z) \cr
-&Z_-(q^{-{k+2 \over 2}}z)
U_-(q^{-{k \over 2}}z)W_-(q^{-{k \over 2}}z)Y^-(z)\biggr\}:.\cr
}
\eqno(17)
$$
where
$$
\eqalign{
Y^+(z)&=\exp\biggl\{-\sum_{m=1}^\infty \qa {z^m \over [2m]}(a_{-m}+\bar
a_{-m})\biggr\}
e^{-{a+\bar a \over 2}}z^{-{a_0+\bar a_0 \over 2}} \cr
&\exp\biggl\{\sum_{m=1}^\infty \qa\qkb {z^{-m} \over [2m]}(a_m+\bar
a_m)\biggr\},  \cr
Y^-(z)&=\exp\biggl\{\sum_{m=1}^\infty \qb {z^m \over [2m]}(a_{-m}+\bar
a_{-m})\biggr\}
e^{a+\bar a \over 2}z^{a_0+\bar a_0 \over 2} \cr
&\exp\biggl\{-\sum_{m=1}^\infty \qb\qkb {z^{-m} \over [2m]}(a_m+\bar
a_m)\biggr\},  \cr
}
\eqno(18)
$$
\noindent
$$
\eqalign{
W_+(z)&=\exp\biggl\{(q-q^{-1})\sum_{m=1}^\infty z^{-m} b_m\biggr\}q^{b_0},  \cr
W_-(z)&=\exp\biggl\{-(q-q^{-1})\sum_{m=1}^\infty q^{-2m}z^{m}
b_{-m}\biggr\}q^{-b_0},   \cr
}
\eqno(19)
$$
\noindent
$$
\eqalign{
Z_+(z)&=\exp\biggl\{-(q-q^{-1})\sum_{m=1}^\infty
z^{-m} {[m] \over [2m]} \bar a_{m})\biggr\}q^{-{1 \over 2}\bar a_0}, \cr
Z_-(z)&=\exp\biggl\{(q-q^{-1})\sum_{m=1}^\infty
\qka z^m{[m] \over [2m]} \bar a_{-m})\biggr\}q^{{1 \over 2}\bar a_0} \cr
},
\eqno(20)
$$
\noindent
$$
\eqalign{
U_+(z)&=\exp\biggl\{(q-q^{-1})\sum_{m=1}^\infty  q^{km}z^{-m}
 {[(k+2)m] \over [2m]}(a_{m}+ \bar a_{m})\biggr\}
q^{{k+2 \over 2}(a_0+\bar a_0)}, \cr
U_-(z)&=\exp\biggl\{-(q-q^{-1})\sum_{m=1}^\infty q^{-2m} z^m
 {[(k+2)m] \over [2m]} (a_{-m}+\bar a_{-m})\biggr\}
q^{-{k+2 \over 2}(a_0+\bar a_0)}. \cr
}
\eqno(21)
$$
\noindent
As we arrange operators in the way of normal ordering
except for the zero modes,
:: denotes normal ordering with respect to the zero modes,
$\alpha<\alpha_0,\bar \alpha<\bar \alpha_0$
and $\beta<\beta_0$.
It is the essential idea of this construction that we define
$q$-deformations of derivative, namely, the field,
$\partial\chi(z)$, $\partial(\chi(z)
-\sqrt{-1}\sigma(z))$ and $\partial \varphi(z)$ as
$$
\eqalign{
&{-2 \over q-q^{-1}}\bigl[Z_+(z)-Z_-(z)\bigr], \cr
&{1 \over q-q^{-1}}\bigl[U_+(z)-U_-(z)\bigr], \cr
&{1 \over q-q^{-1}}\bigl[W_+(z)-W_-(z)\bigr], \cr
}
\eqno(22)
$$
\noindent
respectively
\footnote*{This idea is due to the paper[9]}.
By taking a limit $q \to 1$, they become
$$
\sum_{m \in Z}\bar a_mz^{-m}
,\qquad{k+2 \over 2}\sum_{m\in Z}(a_m+\bar a_m)z^{-m},
\qquad \sum_{m \in Z}b_mz^{-m},
$$
which correspond to the fields
 $z\partial\chi(z), z\partial(\chi(z)
-\sqrt{-1}\sigma(z))$ and $z\partial \varphi(z)$, respectively.
The currents of $U_q(\widehat{\sl}_2)$ satisfy the following relations[6]:
$$
\eqalign{
&K_+(z)X^\pm(w)=\biggl({q^2z-q^{\mp{k \over 2}}w \over z-q^{2\mp{k \over 2}}w}
\biggr)^{\pm 1}X^\pm(w)K_+(z), \cr
&K_-(z)X^\pm(w)=\biggl({q^2w-q^{\mp{k \over 2}}z \over w-q^{2\mp{k \over 2}}z}
\biggr)^{\mp 1}X^\pm(w)K_-(z), \cr
&{q^{2+k}z-w \over q^k z-q^2 w}K_-(z)K_+(w)=
{q^{2-k}z-w \over q^{-k} z-q^2 w}K_+(w)K_-(z), \cr
&X^\pm(z)X^\pm(w)={q^{\pm2}z-w \over z-q^{\pm2}w}X^\pm(w)X^\pm(z), \cr
&X^+(z)X^-(w)\sim {1 \over \qq}\biggl({z \over z-q^kw}K_+(q^{k \over 2}w)
-{z \over z-q^{-k}w}K_-(q^{-{k \over 2}}w)\biggr). \cr
}
\eqno(23)
$$
Here we define the mode expansions of these currents as
$$
\eqalign{
&K_+(z)=\sum_{m\in Z_{\geq 0}}\psi_mz^{-m}=q^{a_0+b_0}\exp\biggl\{
(q+q^{-1})\sum^\infty_{m=1}H_mz^{-m}\biggr\}, \cr
&K_-(z)=\sum_{m\in Z_{\geq 0}}\psi_{-m}z^{m}=q^{-(a_0+b_0)}\exp\biggl\{
-(q+q^{-1})\sum^\infty_{m=1}H_{-m}z^{m}\biggr\}, \cr
&H(z)=\sum_{m\in Z}H_mz^{-m},\qquad X^\pm(z)=\sum_{m\in Z}X^\pm_m z^{-m}. \cr
}
\eqno(24)
$$
Putting $K=q^{a_0+b_0}$ we obtain the relations of the Drinfeld
realization of $U_q(\widehat{\sl}_2)$ for level k[7]:
$$
\eqalign{
&[H_m,H_n]=\delta_{m+n,0}{1 \over m}[2m][km],\qquad m\ne 0,  \cr
&[H_m,K]=0, \cr
&KX^\pm_mK^{-1}=q^{\pm2}X^\pm_m, \cr
&[H_m,X_n^\pm]=\pm{1 \over m}[2m]q^{\mp{k\mid m \mid \over 2}}X^\pm_{m+n},\cr
&X^\pm_{m+1}X^\pm_n-q^{\pm2}X^\pm_nX^\pm_{m+1}
=q^{\pm2}X^\pm_mX^\pm_{n+1}-X^\pm_{n+1}X^\pm_m, \cr
&[X^+_m,X^-_n]={1 \over q-q^{-1}}\biggl(q^{k(m-n) \over 2}\psi_{m+n}
-q^{k(n-m) \over 2}\varphi_{m+n}\biggr). \cr
}
\eqno(25)
$$
The Drinfeld realization of  $U_q(\widehat{\sl}_2)$ corresponds to
one of the $q$-deformation of the algebra $\widehat{\sl}_2$(6).
\par
Next we introduce the Fock module $F_{l,m_1,m_2}
(l\in{1 \over 2}Z;m_1,m_2\in Z)$ freely generated by
$\{a_n,\bar a_n, b_n;n\in Z_{>0}\}$ from a vector
$$
\mid l,m_1,m_2>=\exp\biggl\{l{b \over k+2}+m_1{a \over 2}
-m_2{\bar a \over 2}\biggr\}\mid 0>.
\eqno(26)
$$
\noindent
Here a vector $\mid0>$ has the following properties:
$$
b_n\mid0>=0,\qquad a_n\mid0>=0,\qquad \bar a_n\mid0>=0,
\qquad n\geq0.
$$
The vector $\mid l,m_1,m_2>$ is a eigenvector of $b_0,a_0$
and $\bar a_0$ belonging to eigenvalues $2l, 2m_1$ and
$ 2m_2$, respectively.
{}From the following relations:
$$
[H(z),a_0+\bar a_0]=0,\qquad[X^\pm(z),a_0+\bar a_0]=0,
\eqno(27)
$$
we can restrict the Fock module $F_{l,m_1,m_2}$ to the sector
in which the eigenvalue of $a_0+\bar a_0$ is equal to zero[11].
It is easy to check the vector $\mid{j \over 2},{j \over 2},{j \over 2}>$
satisfies the highest weight conditions.
\beginsection{4. Relations to other realizations}\par
\noindent
Now we will give  relations between our realization of
 $U_q(\widehat{\sl}_2)$  and another one studied by Matsuo.
His realization is based on the following currents of  $\widehat{\sl}_2$[13]:
$$
\eqalign{
&J^\pm(z)=:{1 \over \sqrt{2}}\biggl[\sqrt{k+2}\partial
\phi_1(z)\pm\sqrt{-1}\sqrt{k}\partial\phi_2(z)\biggr]
\exp\biggl\{\pm\sqrt{2 \over k}[\sqrt{-1}\phi_2(z)-\phi_0(z)]\biggr\}:, \cr
&J^3(z)=-\sqrt{k \over 2}\partial\phi_0(z). \cr
}
\eqno(28)
$$
\noindent
We define the following linear transformation of oscillators
$\{a_n,\bar a_n, b_n;n\in Z;a,\bar a, b\}$:
$$
\left(
\matrix{\alpha_m  \cr
        \bar \alpha_m  \cr
         \beta_m    \cr}
\right)=
\left(
\matrix{1      & 0      & 1                            \cr
        -{[(k+2)m] \over [2m]}q^{km}  &  -{[km] \over [2m]}q^{(k+2)m}  & -1
  \cr
         {[(k+2)m] \over [2m]}q^{(k+1)m}  &  {[(k+2)m] \over [2m]}q^{(k+1)m}  &
q^{m} \cr}
\right)
\left(
\matrix{a_m    \cr
        \bar a_m  \cr
        b_m    \cr}
\right),\qquad m\geq 1,
$$
$$
\left(
\matrix{\alpha_{-m}  \cr
        \bar \alpha_{-m}  \cr
         \beta_{-m}    \cr}
\right)=
\left(
\matrix{q^{-(k+2)m}      & 0      & q^{-2m}                            \cr
        -{[(k+2)m] \over [2m]}q^{-2m}  &  -{[km] \over [2m]}  & -q^{-2m}
\cr
         {[(k+2)m] \over [2m]}q^{-m}  &  {[(k+2)m] \over [2m]}q^{-m}  & q^{-m}
\cr}
\right)
\left(
\matrix{a_{-m}    \cr
        \bar a_{-m}  \cr
        b_{-m}    \cr}
\right),\qquad m\geq
 1,
$$
$$
\left(
\matrix{\alpha_{0}  \cr
        \bar \alpha_{0}  \cr
         \beta_{0}    \cr}
\right)=
\left(
\matrix{1 & 0 & 1 \cr
        -{k+2 \over 2} & -{k \over 2} & -1  \cr
         {k+2 \over 2} & {k+2 \over 2} & 1   \cr}
\right)
\left(
\matrix{a_0    \cr
        \bar a_0  \cr
        b_0    \cr}
\right).
$$
$$
\left(
\matrix{\alpha  \cr
        \bar \alpha  \cr
         \beta    \cr}
\right)=
\left(
\matrix{1 & 0 & 1 \cr
        -{k+2 \over 2} & -{k \over 2} & -1  \cr
         {k+2 \over 2} & {k+2 \over 2} & 1   \cr}
\right)
\left(
\matrix{a    \cr
        \bar a  \cr
        b    \cr}
\right).
\eqno(29)
$$
{}From the commutation relations(16) one can obtain
$$
\eqalign{
&[\alpha_m,\alpha_n]=\delta_{m+n,0}{[2m][km]\over m},\qquad
[\alpha_0,\alpha]=2k, \cr
&[\bar \alpha_m,\bar \alpha_n]=-\delta_{m+n,0}{[2m][km]\over m},
\qquad [\bar \alpha_0,\bar \alpha]=-2k, \qquad m,n\not=0 \cr
&[\beta_m,\beta_n]=\delta_{m+n,0}{[2m][(k+2)m]\over m},\qquad
[\beta_0,\beta]
=2(k+2). \cr
}
\eqno(30)
$$
Substituting the inverse transformations of (29) into currents(17),
they become
$$
\eqalign{
K_+(x)=&\exp\biggl\{(q-q^{-1})\sum_{m=1}^{\infty}z^{-m}\alpha_m\biggr\}
q^{\alpha_0},  \cr
K_-(x)=&\exp\biggl\{-(q-q^{-1})\sum_{m=1}^{\infty}z^{m}
\alpha_{-m}\biggr\}
q^{-\alpha_0},  \cr
X^+(z)=&{1 \over q-q^{-1}}:\biggl\{Y^+(z)Z_+(q^{-{k+2 \over 2}}z)
W_+(q^{-{k \over 2}}z) \cr
&-W_-(q^{{k \over 2}}z)
Z_-(q^{{k+2 \over 2}}z)Y^+(z)\biggr\}:,    \cr
X^-(z)=&-{1 \over q-q^{-1}}:\biggl\{Y^-(z)Z_+(q^{{k+2 \over 2}}z)
W_+(q^{k \over 2}z)^{-1} \cr
&-W_-(q^{-{k \over 2}}z)^{-1}Z_-(q^{-{k+2 \over 2}}z)
Y^-(z)\biggr\}:,\cr
}
\eqno(31)
$$
where
$$
\eqalign{
Y^+(z)&=\exp\biggl\{\sum_{m=1}^\infty \qa {z^m \over [km]}
(\alpha_{-m}+\bar \alpha_{-m})\biggr\}
e^{\alpha+\bar \alpha \over k}z^{{\alpha_0+\bar \alpha_0 \over k}} \cr
&\exp\biggl\{-\sum_{m=1}^\infty \qa {z^{-m} \over [km]}
(\alpha_m+\bar \alpha_m)\biggr\},  \cr
Y^-(z)&=\exp\biggl\{-\sum_{m=1}^\infty \qb {z^m \over [km]}
(\alpha_{-m}+\bar \alpha_{-m})\biggr\}
e^{-{\alpha+\bar \alpha \over k}}z^{-{\alpha_0+\bar \alpha_0 \over k}} \cr
&\exp\biggl\{\sum_{m=1}^\infty \qb {z^{-m} \over [km]}
(\alpha_m+\bar \alpha_m)\biggr\},  \cr
}
\eqno(32)
$$
\noindent
$$
\eqalign{
W_+(z)&=\exp\biggl\{-(q-q^{-1})\sum_{m=1}^\infty z^{-m}
{[m] \over [2m]}\beta_m\biggr\}q^{-{\beta_0 \over 2}},   \cr
W_-(z)&=\exp\biggl\{(q-q^{-1})\sum_{m=1}^\infty z^{m}
{[m] \over [2m]}\beta_{-m}\biggr\}q^{\beta_0 \over 2},   \cr
}
\eqno(33)
$$
\noindent
$$
\eqalign{
Z_+(z)&=\exp\biggl\{-(q-q^{-1})\sum_{m=1}^\infty
z^{-m} {[m] \over [2m]} \bar \alpha_{m})\biggr\}q^{-{1 \over 2}\bar \alpha_0},
 \cr
Z_-(z)&=\exp\biggl\{(q-q^{-1})\sum_{m=1}^\infty
z^m{[m] \over [2m]} \bar \alpha_{-m})\biggr\}q^{{1 \over 2}\bar \alpha_0}. \cr
}
\eqno(34)
$$
These currents correspond with those in the paper[9] except for
a little change
because of a normalization
of zero mode.
\par
Shiraishi's representation is also based on the Wakimoto currents.
The main difference is a treatment for $q$-deformation of derivative.
He extends derivative to a $q$-difference operator defined as
$$
\;_n\partial_z f(z)\equiv{f(q^nz)-f(q^{-n}z) \over (q-q^{-1})z}.
\eqno(35)
$$
It is not easy to show explicitly relations between his representation
and ours. However, there is a correspondence between $H(z)$,
$X^\pm(z)=X^\pm_+(z)+X^\pm_-(z)$ in (17)(24) and $zJ^3(z)$,
$zJ^\pm(z)=J^\pm_I(z)+J^\pm_{I\hskip -2pt I}$
in the appendix of the paper[14].
\par
Finally the aim of this paper is concentrated on the forms of currents
and relations of them. In the line of our construction
we can obtain screening currents, qVOs
and n-point correlation functions with one screening charge on sphere.
In the case of the two-point correlation function,
we confirmed correspondence with the results of the paper[14].
Our results will be contained in a forthcoming paper. It is necessary to
investigate cohomological structure of the Fock module[15][16] in order
to derive irreducible representations and correlation functions
on torus. These are under investigation.
\beginsection{Acknowledgment}\par
\noindent
The author thanks A.Matsuo for helpful discussions and teaching me his works.
\vbn
\beginsection{References}\par
\medn
\item{[1]}  M.\ Wakimoto,
Commun.\ Math.\ Phys.\ {\bf 104} (1986) 605.
\item{[2]}  I.B.\ Frenkel and N.Yu.\ Reshetikhin, Commun.\ Math.\ Phys.\
{\bf 146} (1992) 1.
\item{[3]} B.\ Davies, O.\ Foda, M.\ Jimbo, T.\ Miwa, A.\ Nakayashiki,
Diagonalization of the XXZ Hamiltonian by vertex operators. preprint RIMS-873
(1992) to appear in Comm.\ Math.\ Phys..
\item{[4]} M.\ Jimbo, K.\ Miki, T.\ Miwa and A.\ Nakayashiki,
Phys.\ Lett.\ {\bf A168} (1992) 256.
\item{[5]} M.\ Idzumi, K.\ Iohara, M.\ Jimbo, T.\ Miwa, T.\ Nakashima,
T.\ Tokihiro, Quantum Affine Symmetry in vertex models, RIMS preprint (1992)
to appear in Int.\ J.\ Mod.\ Phys.\ A.
\item{[6]} I.B.\ Frenkel and N.H.\ Jing, Proc.\ Nat'l Acad.\ Sci.\ USA {\bf85}
(1988) 9373.
\item{[7]} V.G.\ Drinfeld,\ Sov.\ Math.\ Dokl.\ {\bf 36} (1987) 212.
\item{[8]} J.\ Shiraishi.\ Free boson representation of
$U_q(\widehat{\sl}_2)$, preprint UT-617 (1992).
\item{[9]} A.\ Matsuo, Free field Represetation of Quantum Affine Algebra
$U_q(\widehat{\sl}_2)$, preprint (1992).
\item{[10]} A.\ Abada, A.H.\ Bougourzi, M.A.\ El\ Gradechi,
Deformation of the Wakimoto construction, preprint (1992).
\item{[11]}  D.\ Friedan, E.\ Martinec, S.\ Shenker,
Nucl.\ Phys.\ {\bf B271} (1986) 93.
\item{[12]} A.Gerasimov, A.Marshakov, A.Morozov, M.Olshanetsky
and S.Shatashvili, Int.\ J.\ Mod.\ Phys.\ {\bf A5} (1992) 2495.
\item{[13]}  D.\ Nemeschansky, Phys.\ Lett.\ {\bf B224} (1989) 121.
\item{[14]} A.\ Kato, Y.\ Quano, J.\ Shiraishi, Free boson representation of
$q$-vertex operators and their correlation functions preprint UT-618 (1992).
\item{[15]} D.\ Bernard and G.\ Felder, Commun.\ Math.\ Phys.\
{\bf 127} (1990) 145.
\item{[16]} B.\ Feigin and E.\ Frenkel, Commun.\ Math.\ Phys.\ {\bf 128} (1990)
161.
\end